\documentstyle[12pt, epsf] {article}
\begin{document}

\title{Physical and Geometric\\ Interpretations of the \\
 Riemann Tensor, Ricci Tensor, \\ and Scalar Curvature}
\author{Lee C. Loveridge}
\date {\today}
\maketitle

\begin{abstract}
Various interpretations of the Riemann Curvature Tensor, Ricci Tensor, 
and Scalar Curvature are described.  Also the physical meanings of the 
Einstein Tensor and Einstein's Equations are discussed.  Finally, a derivation
of Newtonian Gravity from Einstein's Equations is given.
\end{abstract}

\section{Introduction}
Electromagnetism is taught at a variety of levels with various combinations of
intuitive pictures and hard equations.  One generally begins (often in 
elementary school) with such simple concepts as opposites attract, rubbing
a balloon on your hair makes it charged, and magnets have two distinct poles.
Later Coulomb's Law may be introduced and then the integral forms of Maxwell's
equations which have very intuitive pictoral representations.  Charges are the
sources (and sinks) for all electric charge (Gauss' Law of Electricity)
Currents cause circulation of magnetic fields (Ampere's Law).  Later the 
derivative forms and covarient forms of Maxwell's equations can be added 
building upon these foundations.

Unfortunately, such is generally not the case in teaching General Relativity.
Either one is presented with a very vague elementary presentation about space
being like a globe where innitially parallel lines end up crossing, or one gets
the full graduate treatment, based on complicated indeces or advanced
differential geometry.  There is very little middle ground, and very little 
intuitive explanation at the advanced level.  

I believe that part of the reason for this is a lack of well known intuitive
geometrical explanations of the important objects in GR such as the 
Ricci Tensor, Scalar Curvature, and Einstein's Tensor.  Such pictures would
make it easier to visualize the equations of GR and allow teaching the
concepts at a middle level.  

Parallel transport is actually fairly well described in most elementary 
discussions so I will not focus on it.  The Riemann Tensor is also often well 
described in a geometrical sense in more advanced discussions.  One could just 
say that the Ricci Tensor, Scalar Curvature, and Einstein's Tensor are just 
specific linear combinations of the Riemann Tensor.  One could also say that 
divergences and curls are just linear combinations of partial derivatives
on vector fields.  That is true, but overlooks the vital facts, that divergence
represents the overall increase in flux density, and curl represents the amount
of circulation.  Both of which are vitally important ideas for a conceptual 
understanding.  

Rather than trying to make a self contained mid-level treatment of GR,
this article is written for those like myself who have graduate training in 
the field, but seek to better understand the fundamental objects so they can 
better explain them to others.  However, the main body of the text should
be understandable to any with an intuitive understanding of parallel transport
and geodesics.

\section{Conventions}
I will begin by reviewing my conventions, which in general follow reference
\cite{Nakahara}.

Throughout this paper I will restrict myself to torsion free spaces with 
a metric and a metric-connection, 
$$\Gamma^{\kappa}_{\hspace{5 pt} \mu \nu} =
\frac 1 2 g^{\kappa \lambda}(\partial_{\mu}g_{\lambda \nu}
    +\partial_{\nu}g_{\lambda \mu}- \partial_{\lambda}g_{\mu \nu}).$$
The Riemann curvature tensor is then
$$R^{\kappa}_{\hspace {5 pt} \lambda \mu \nu}=
	 \partial_{\mu} \Gamma^{\kappa}_{\hspace{5 pt} \nu \lambda}
	-\partial_{\nu} \Gamma^{\kappa}_{\hspace{5 pt} \mu \lambda}
	+\Gamma^{\eta}_{\hspace{5 pt} \nu \lambda}
		\Gamma^{\kappa}_{\hspace{5 pt} \mu \eta}
	-\Gamma^{\eta}_{\hspace{5 pt} \mu \lambda}
		\Gamma^{\kappa}_{\hspace{5 pt} \nu \eta}.$$

Vectors will be represented with bold 
faced roman letters.  Vector components will be represented with superscripts.
I will also use the following symbols:
$$({\bf T}, {\bf S})=g_{\mu \nu}T^{\mu}T^{\nu}$$
$$R({\bf P, Q, S, T})=
	R_{\mu \nu \rho \sigma} P^{\mu}Q^{\nu}S^{\rho}T^{\sigma}.$$

The Ricci tensor is the contraction of the Riemann tensor, and will be written
as $R$ with just two indeces or two arguments
$$R_{\mu \nu}=R^{\rho}_{\hspace{5 pt} \mu \rho \nu}.$$
The scalar curvature is the contraction of the Ricci tensor, and is written
as $R$ without subscripts or arguments
$$R=g^{\mu \nu}R_{\mu \nu}.$$

Often times, partial derivatives will be represented with a comma
$$\partial_\mu A=A,_\mu.$$
 
\section{Gaussian Curvature of a Two-Dimensional Surface}
I will begin by describing Gauss' notion of internal curvature.  I will 
basically follow reference \cite{MKW}.  Prior to Gauss people had studied 
the notion of extrinsic curvature.  That is, they studied the way a curve
imbedded in a two dimensional space turned as one moved along it, and at any 
given point they could find the 
effective radius $\rho$ of a circle that would turn at the same rate as the curve at 
that point.  The curvature was then simply the inverse of this effective 
radius. $\kappa = 1/\rho$.  A two dimensional surface imbedded in a three
dimensional space could be described by two such radii or curvatures.
In 1827, Gauss had an idea for defining curvature 
in an intrinsic way, that is in a way that could be measured without embedding
 the surface in a higher dimensional space.  Imagine starting at a point in 
a space and moving a geodesic distance $\epsilon$ in all directions.  In 
essence you would form a the equivalent of a ``circle'' in this space.  If
the space is flat, then the circumference $C$ of this circle will simply be 
$C=2 \pi \epsilon$ but in a curved space the circumference will be slightly 
greater or smaller than this.  For example on a sphere of radius $\rho$, the
circumference would be 
$$C=2 \pi \rho \sin{(\epsilon/\rho)} \approx 
	2 \pi \epsilon(1 - \frac {{\epsilon}^2} {6 {\rho}^2})$$
For more complicated surfaces the expression is not so simple, but in general
the lowest order deviation will be quadratic as in this case.  Gauss, then
suggested the following definition for curvature $\kappa_{Gauss}$.
\begin{equation}
\kappa_{Gauss}=\lim_{\epsilon \to 0} {\frac {6} {{\epsilon}^2} \Bigl( 1 -
	\frac {C}{2 \pi \epsilon} \Bigr)}.
\end{equation}
It turns out that this intrinsic curvature is just the product of the two
extrinsic curvatures of the surface.  

What is most important for this paper 
however, is that we have an intrinsic measure of curvature with a clear 
and concrete physical/geometrical meaning.  The Gaussian Curvature 
$\kappa_{Gauss}$ is a measure of how much the circumference of a small
circle deviates
from its expected value in a flat space.  The deviation is quadratic, as 
we would expect since our usual measure of curvature, the Riemann tensor, 
depends on second derivatives of the metric.  We will find that all of our
physical interpretations of curvature will depend on quadratic variations. 
Constant and linear variations are simply artifacts of the coordinate system.

There are a number of other relationships between the gaussian curvature and
physical quantities.  However, they are well detailed in the literature
\cite{MKW} and this one interpretation is sufficient for this paper. 

\section{Interpretations of the Rieman Tensor}
	There are a variety of physical/geometrical interpretations of the
Riemann Tensor.  Most texts on gravitation or Riemannian geometry will 
present at least
one of them.  Therefore I will present them only briefly.

\subsection{Collection of Gaussian Curvatures}
	I believe this interpretation is popular among mathematicians, and 
can be found in \cite{MKW}.  If you have two non-parallel vectors $S$ and $T$
at a point in your space, then they define a two dimmensional subspace.  
The quantity 
$$R_{\mu \nu \rho \sigma} S^{\mu} T^{\nu} S^{\rho} T^{\sigma}$$
is simply the gaussian curvature of that subspace times the area squared 
of the ${\bf S}$ ${\bf T}$ parallelogram.
In other words
\begin{equation}
\kappa_{G,T \wedge S} \left| \begin{array} {cc}
				({\bf S,S}) & ({\bf S,T}) \\
			({\bf T,S}) & ({\bf T,T}) \end{array} \right| 
= R({\bf S,T,S,T}).
\label{RiemannasGauss} 
\end{equation}

This equation can be proven by direct calculation of the Gaussian Curvature
as described above.  (Define a ``circle'', find its circumference, and find
its deviation from the circumference in flat space.  This method is followed in
depth for $D$ dimensions in appendix \ref{Scalarproof}.  Here only two
dimensions are required.)  The calculation is 
straight forward though tedious, and care must be taken to include all 
second order variations, including changes in the metric.  The
area dependence is most easily found by first allowing ${\bf S}$ and 
${\bf T}$ to be 
orthonormal, and then generalizing.   

\subsection{Deviation of a Vector}
This is perhaps the most common interpretation used in physics texts and can
be found in \cite{Nakahara}, \cite{MKW}, or \cite{Wald}.  Begin with a 
vector $A$ and parallel transport it around a small parallelogram defined by
the vectors ${\bf u}$ and ${\bf v}$ (starting with diplacement along 
${\bf u}$).  The deviation of the vector ${\bf A}$ can then be found as
\begin{equation}
\delta A^{\alpha} = - R^{\alpha}_{\hspace{5 pt} 
			\beta \mu \nu} A^{\beta}u^{\mu}v^{\nu}. 
\end{equation}
\label{vecdev}
This again can be shown by straight forward calculation carrying all terms
to second order.

\subsection{Geodesic Deviation Equation} 
This is another common interpretation used in physics texts, including
\cite{Wald} that is very 
useful for understanding General Relativity.  Consider a one parameter
family of geodesics $x^{\mu}(s,t)$ where $s$ is the parameter and $t$ the 
geodesic flow.  $S^{\mu}=dx^{\mu}/ds$ is the vector describing motion 
from one geodesic to another and $T^{\mu}=dx^{\mu}/dt$ describes flow along 
the geodesic.  Then we find that if ${\bf S}$ is parallel transported along 
the geodesic flow, its second covariant derivative is determined by the
Riemann tensor:
\begin{equation}
\frac {D^2 S^{\mu}} {dt^2} = -R^{\mu}_{\hspace{5 pt}
					\nu \sigma \rho} S^{\sigma} 
				T^{\nu} T^{\rho}.
\label{geodev}
\end{equation}

While the description of this equation may seem forboding, its meaning can be
made more physically clear (if somewhat less technically accurate) by 
imagining that ${\bf S}$ represents the separation between two objects
 near each other in space and ${\bf T}$ represents their innitial motion.
The equation then simply describes their relative acceleration.

\section{Interpretations of the Ricci Tensor}
We now turn our attention to the Ricci Tensor.  The ricci tensor is simply a
contraction of the Rieman tensor, defined as
\begin{equation}
R_{\mu \nu}=R^{\rho}_{\hspace{5 pt}\mu \rho \nu}.
\end{equation}
Thus, if we know all of the Riemann 
tensor we can calculate the Ricci tensor, but we haven't really addressed the
issue of its meaning.  (In the same sense that saying $\nabla \dot{} {\bf E}$ 
is just a linear combination of $\partial_i E_j$ does not explain that the
gradiant must equal the sources of the field.)

\subsection{Sum of Gaussian Curvatures}
Whenever I've asked a mathematician what the Ricci tensor means, they've 
explained the meaning of the Riemann tensor as a collection of gaussian 
curvatures and simply stated that the Ricci tensor was an average.  While I
do not find this explanation very satisfying, it bears further investigation.
Suppose you wanted to find the average curvature of all planes involving
the vector ${\bf S}$.  You could start by taking a collection of orthonormal
 vectors ${\bf t}_i$ and saying
\begin{equation}
\bar{\kappa}_{\bf S}=\frac {1} {D} \sum_{i=1}^D R({\bf S, t_i, S, t_i})
 =\frac {1} {D} R({\bf S,S}).
\end{equation}
We should note that these are actually area weighted curvatures.  That is 
they contain a factor of the area squared of the parallelogram formed by 
${\bf S}$ and ${\bf t_i}$, as in equation \ref{RiemannasGauss}.  
 
One may ask whether summing over one orthonormal set is sufficient.  Perhaps 
we should average over all such orthonormal sets.  The 
Ricci tensor does not depend on your basis, so no further
averaging is required.  So while the Riemann tensor told us the gaussian 
curvature of any given sub plane, the Ricci tensor gives us the average of all
sub planes involving a given vector.

\subsection{Volume Deviation}
I find the previous description lacking because instead
of describing the behavior of a single physical object, it describes an 
average of the behavior of several objects.  There is a way to describe the
physical meaning of the Ricci tensor without invoking the notion of an 
average.  I found the idea for this description in \cite{baez}, but proofs and
full details were not included.

Suppose instead of looking at two small objects in space, we considered a 
volume filling collection of small objects in space.  Describing the relative
acceleration of any two of them would require the geodesic deviation equation,
but to describe the evolution of their volume, we would have to average over
several different versions of the equation.  These have roughly the result of
averaging the Riemann tensor into a Ricci tensor.  
	So in roughly the same 
sense that the Riemann tensor governs the evolution of a vector or a 
displacement parallel propagated along a geodesic, the Ricci tensor governs
the evolution of a small volume parallel propagated along a geodesic.  We
must be careful though.  Unlike vectors, volumes may change along geodesics 
even in a flat space.  We must therefore subtract off any change that would 
occur in a flat space.  Suppose then that we have a small volume 
$\delta V$ of dust near a point $x_0^{\mu}$.  If we allow that volume to move
along a direction $T^{\mu} = \frac {dx^{\mu}} {d\tau}$ we find the following
equation:

\begin{equation}
\frac {D^2} {{d\tau}^2} {\delta}V 
 	- \frac {D_{\rm{flat}}^2} {{d\tau}^2} {\delta}V
	= -{\delta}V R_{\mu \nu} T^{\mu} T^{\nu}. 
\label{riccidef}
\end{equation}

The proof of this equation can be found in appendix \ref{Ricciproof}. 

One may ask whether this volume should be a $D$-dimensional volume (in 
relativity a space time volume) or a $D-1$-dimensional volume (a space only
volume).  It turns out that it can be either, so long as the $D-1$-dimensional
volume is transverse to the vector ${\bf T}$.  Thus, we may apply the equation
to the deviation of a spacelike volume as it propagates through time.

\section{Interpretation of the Scalar Curvature}
	We turn now to the scalar curvature, which is just the contraction of
the Ricci tensor.  It turns out the scalar curvature has a meaning very 
similar to the gaussian curvature.  If we imagine instead of taking a circle,
taking a generalized $D-1$ sphere, i.e. the set of all points a geodesic 
distance $\epsilon$ from a given starting point $x_0^{\mu}$.  We can
calculate the area of this sphere in flat space, but in curved space the 
area will deviate from the one we calculated by an amount proportional to 
the curvature.  Thus, we find that the scalar curvature is
\begin{equation}
R = \lim_{\epsilon \to 0} \frac {6 D} { \epsilon^2} \biggl{[} 1 -
	\frac {A_{\rm{curved}}(\epsilon)} {A_{\rm{flat}}(\epsilon)} 
	\biggr{]}.
\label{scalardef1}
\end{equation}
The proof of this equation is in appendix \ref{Scalarproof}.

Feynman \cite{Feynman1} mentions this definition of curvature in the case of 
three space dimensions.  His, definition differs from the one given here by 
a factor of $-2$ which will be explained in the next section.

\section{The Einstein Tensor}
We now have a physical interpretation for each part of Einstein's equation,
and we could begin immediately to discuss the equation's meaning.  However, 
I'd like to first discuss the particular combination of the Ricci tensor and
scalar curvature known as Einstein's tensor.  
$$G_{\mu\nu}=R_{\mu\nu}-\frac 1 2 R g_{\mu\nu}$$
Feynman \cite{Feynman1} 
discusses the meaning of this tensor to some extent, and I will elaborate on
his ideas, though my presentation will be very different.

Suppose we know the Riemann tensor for some point in a $D$ dimensional space,
but that we wanted to find the scalar curvature not for the full space,
but for the $D-1$ subspace that is orthogonal to a particular vector $\bf t$.
In the specific case of General Relativity in four dimensions, $\bf t$ could
represent the direction of time and we would then be finding the curvature
of the three coresponding spacial dimensions.  We would have to contract the
Riemann tensor, not with $g^{\mu\nu}$ but rather with the projector onto our
$D-1$ space
$$g^{\mu\nu}-t^\mu t^\nu.$$
(Here it is assumed that $\bf t$ is normalized to $1$.)
The $D-1$ dimmensional curvature is then
\begin{equation}
R_{D-1}=\bigl(g^{\mu\nu}-t^\mu t^\nu \bigr)
	\bigl(g^{\alpha\beta}-t^\alpha t^\beta \bigr)
	R_{\mu\alpha,\nu\beta}
       =R-2R_{\mu\nu}t^\mu t^\nu=-2G_{\mu\nu}t^\mu t^\nu.
\label{Einsteintensor}
\end{equation}
So we see that the curvature of the $D-1$ dimmensional subspace orthogonal
to the unit vector $\bf t$ is just
negative two times the Einstein tensor fully contracted with the vector
$\bf t$. For general relativity, this means that once we choose a time 
direction, Einstein's tensor tells us the scalar curvature of the 
corresponding spacial dimensions.

In Feynman's treatment, he apparently includes the factor of $-2$ in his
deffinition of the scalar curvature.  This accounts for the difference in our
definition of scalar curvature in three dimensional space.  The numerical
factor is unimportant so long as it is treated correctly.  The sign change 
compensates for the fact that the spatial dimmensions have a negative metric.
Feynman's sign convention gives the curvature that would be found if we 
treated only the spatial dimensions with a positive metric.  I will continue
with my sign convention and make comments where necesary.

We can now discuss the meaning of Eintein's Equation which reads
$$
G_{\mu\nu}=8 \pi G T_{\mu\nu},
$$
where $T_{\mu\nu}$ is the stress-energy tensor, and $G$ is the Universal
Gravitational Constant.  When the stress-energy
tensor is contracted with a timelike unit vector $\bf t$ we get the energy 
density $\rho$ in the 
frame of reference defined by $\bf t$.  Thus If we contract Einstein's
equation with the vector $\bf t$ giving the time direction,  we get
\begin{equation}
 R_3 = -4 \pi G \rho.
\end{equation}
Thus, the scalar curvature of the spatial dimensions equal $-4 \pi G$ times
the energy density in any chosen frame of reference.  This is the meaning of 
the Einstein equation.  Curvature equals energy density, and the equality 
must hold in all frames.  

The sign may appear misleading, as it suggests that
positive energy would give a negative curvature.  However, this is because 
the spatial dimensions have negative signature so that closed surfaces such 
as spheres have negative curvature. Because of Einstein's equation we learn
that the earth for example does not actually have the shape of a ball in
flat space, but of a ball at one point on a very large three sphere.  However,
the curvature is very small.  As Feynmann points out, if we calculated the 
change in radius due to this curvature we would find that it is approximately
1 fermi for every 4 billion metric tons. 

\section{Deriving Newtonian Gravity}
Now let's try to use our results to derive Newtonian gravity.  We will 
restrict ourselves to 4 dimensions (3 space, 1 time).  First we begin
with Einstein's equation written in terms of the Ricci tensor and scalar
curvature
\begin{equation}
R_{\mu\nu}-\frac 1 2 R g_{\mu\nu}=8 \pi G T_{\mu\nu}
\label{einstein2}
\end{equation}
and contract both sides to get 
$$
R=-8 \pi G T.
$$ 
Substituting back into equation \ref{einstein2} and rearranging we get
\begin{equation}
R_{\mu\nu} = 8 \pi G \bigl( T_{\mu\nu}-\frac 1 2 T g_{\mu\nu} \bigr).
\label{einstein3}
\end{equation}

Next we fully contract the equation with the timelike unit vector $\bf t$ 
which describes the innitial motion of the matter and get
$$R_{\mu\nu}t^{\mu}t^{\nu} = 8 \pi G (\rho-\frac 1 2 T).$$
If we specialize to a region with matter dominated energy, then $T$ is simply
the energy density in the matter's rest frame $\rho_0$.  
If we allow $\bf t$ to 
describe that same rest frame (or one moving very slowly relative to it), 
then both terms on the right hand side contain the same factor of $\rho$ so
that we get,
$$  R_{\mu\nu}t^{\mu}t^{\nu} = 4 \pi G \rho. $$

Next we substitute this into equation \ref{riccidef} and get
$$
\frac {D^2} {{d\tau}^2} {\delta}V 
 	- \frac {D_{\rm{flat}}^2} {{d\tau}^2} {\delta}V
	=4 \pi G \rho {\delta}V.
$$
We integrate this over space to put it in terms of a finite spherical volume
of radius $r$ so that our equation is
$$
\frac {D^2} {{d\tau}^2} V 
 	- \frac {D_{\rm{flat}}^2} {{d\tau}^2} V
	=4 \pi G M_{enclosed}.
$$

Now let's consider the time derivative.  As long as the objects experiencing
the gravitational force are moving slowly, and the force is small, the
covariant derivative is just a regular derivative with respect to time.  Thus,
we get
$$
\ddot{V} = 4 \pi r^2 \ddot{r} + 8 \pi r {\dot{r}}^2.
$$
The second term on the left is the one which would be there even in flat space
(no gravity)  The first term is then the one which enters into our equation 
and we get
$$4 \pi r^2 \ddot{r}=-4 \pi G M_{enc},$$
or
\begin{equation}
\ddot{r}=-\frac {G M_{enc}} {r^2}.
\end{equation}
This is of course Newton's equation for universal gravity (for the specific
case of slow moving matter).  It should be noted that by using the standard 
formula for the volume of the sphere we were already ignoring higher order
corrections.  

It is interesting to note that if we contract equation \ref{einstein3}
with a lightlike vector  $\bf t$ instead of a timelike one, then the trace
term $T$ drops out and we get an acceleration that is twice as big.  This may
explain why when describing the bending of light, General Relativity gives a
result that is twice as big as the one from Newtonian Gravity.

\section{Conclusion}
By now we have looked in depth at the various important curvature 
quantities in General Relativity.  We have studied their geometric and 
physical meanings, and we have seen how the lowest order Newtonian 
Gravitation is a result of Einstein's equation and the curvature of space.  
I hope this is helpful for your own understanding of General Relativity and in
teaching it.

\appendix

\section{Proof of Volume Deviation Equation }
\label{Ricciproof}

We begin by defining a small volume described by a set of $D$ vectors 
$\{ {\bf t_i} \}$, where $D$ is 
the dimension of the space.  The vectors must be linearly independent, 
otherwise they would not span a $D$-volume.  The volume can then be written 
as 
\begin{equation}
\delta V = \prod_{i=1}^{D} t_i^{\mu_i} 
	\sqrt{g}\epsilon_{{\mu_1}{\mu_2}{\ldots}
{\mu_D}} = {\rm{det}}(t_i^{\mu}) = \prod_{\mu=1}^{D} t_{\alpha_\mu}^{\mu} 
\sqrt{g}\epsilon^{{\alpha_1}{\alpha_2}{\ldots}{\alpha_D}}.
\end{equation}

The tensor $\sqrt{g}\epsilon_{{\mu_1}{\mu_2}{\ldots}{\mu_D}}$ is covariantly
invariant, so when we take covariant derivatives of the volume we need only 
act on the vectors.  We can then take covariant derivatives along a curve 
described by the vector ${\bf T}$ and the parameter $\tau$.  The first 
derivative is
$$\frac {D} {d\tau} \delta V = \sum_{ \mu=1}^{D} \dot{t}^{\mu}_{\alpha_\mu} 
\prod_{\nu=1 \atop \nu \neq \mu}^{D} t_{\alpha_\nu}^{\nu} 
\sqrt{g}\epsilon^{{\alpha_1}{\alpha_2}{\ldots}{\alpha_D}},$$
and the second derivative is
\begin{equation}
\frac {D^2} {{d \tau}^2} \delta V = \sum_{ \mu=1}^{D} 
	\ddot{t}^{\mu}_{\alpha_\mu} 
\prod_{\nu=1 \atop \nu \neq \mu}^{D} t_{\alpha_\nu}^{\nu} 
\sqrt{g}\epsilon^{{\alpha_1}{\alpha_2}{\ldots}{\alpha_D}} + 
\sum_{ \mu,\nu=1\atop \nu \neq \mu}^{D} \dot{t}^{\mu}_{\alpha_\mu} 
	\dot{t}^{\nu}_{\alpha_\nu} 
\prod_{\rho=1 \atop \rho \neq \mu, \nu}^{D} t_{\alpha_\rho}^{\rho} 
\sqrt{g}\epsilon^{{\alpha_1}{\alpha_2}{\ldots}{\alpha_D}}.
\label{Vaccel}
\end{equation}
In these equations 
$$\dot{t}^\mu_{\alpha_\mu}=\frac {D} {d \tau} t_{\alpha_\mu} = 
 \nabla_T t^\mu_{\alpha_\mu},$$
 and 
$$\ddot{t}_{\alpha_\mu}=\frac {D^2} {{d \tau}^2} t^\mu_{\alpha_\mu} = 
 \nabla_T (\nabla_T t^\mu_{\alpha_\mu})=
-R^\mu_{{\hspace {4 pt} }{\nu}{\rho}{\sigma}}t^{\rho}_{\alpha_\mu}
	T^{\nu}T^{\sigma}$$

The second term in equation \ref{Vaccel} would appear even if the space were 
flat.  It simply refers to the expanding or contracting of the volume in flat 
space.  For example in Euclidean three space we can imagine a volume of 
constant solid
angle $d \Omega$ and constant radial width $dr$ moving radially with constant
radial velocity.  All points in the volume are moving along geodesics, the 
space is clearly flat, but the volume is increasing quadratically with time,
and therefore has a non-zero second derivative.  I do wonder if a rescaling 
of $\tau$ couldn't set this to zero, but at this point I'm uncertain.  This 
is the term refered to as $\frac {D_{\rm{flat}}^2} {{d\tau}^2} {\delta}V$ in 
equation \ref{riccidef}.  It is subtracted off in that equation, and plays
no further role here. 

Substituting the geodesic deviation equation \ref{geodev} into the first term 
in equation \ref{Vaccel} we get

$$  
\sum_{ \mu=1}^{D} \ddot{t}^{\mu}_{\alpha_\mu} 
\prod_{\nu=1 \atop \nu \neq \mu}^{D} t_{\alpha_\nu}^{\nu} 
\sqrt{g}\epsilon^{{\alpha_1}{\alpha_2}{\ldots}{\alpha_D}}
= \sum_{ \mu=1}^{D} 
	- R^\mu_{{\hspace {4 pt} }{\phi}{\rho}{\sigma}}t^{\rho}_{\alpha_\mu}
	T^{\phi}T^{\sigma}
\prod_{\nu=1 \atop \nu \neq \mu}^{D} t_{\alpha_\nu}^{\nu} 
\sqrt{g}\epsilon^{{\alpha_1}{\alpha_2}{\ldots}{\alpha_D}}. $$
From here we can see that if $\rho$ is equal to any of the $\nu$'s then we 
are taking the determinent of a matrix with a repeated row and the term is 
therefore $0$.  
We can therefore set $\rho = \mu$ because $\mu$ is the only value not already
appearing in the product.  The sum then factors out and we have
\begin{equation}
(\sum_{ \mu=1}^{D} 
	 -R^\mu_{{\hspace {4 pt} }{\phi}{\mu}{\sigma}})
	(T^{\phi}T^{\sigma})
(\prod_{\nu=1}^{D} t_{\alpha_\nu}^{\nu} 
\sqrt{g}\epsilon^{{\alpha_1}{\alpha_2}{\ldots}{\alpha_D}})=
	-R_{{\phi}{\sigma}} T^{\phi}T^{\sigma} \delta V.
\end{equation}
Thus proving the guess for the Ricci tensor.  

For General Relativity, we would like to think of the parameter $\tau$ as 
representing time, and ${\bf T}$ as the direction of time flow.  We would then
generally want to know the evolution of a space volume not a full space-time 
volume.  Thus, we want to know if the above equation still applies if we deal
with a $D-1$ volume.  The answer is that a $D-1$ volume is allowed so long as
it is transverse to the vector ${\bf T}$.  Because our set of vectors 
$\{ {\bf t_i} \}$ spans space(-time) it must be linearly dependent with 
${\bf T}$.  We can therefore drop one of the ${\bf t_i}$'s in favor of 
${\bf T}$ without changing the volume.  However, the geodesic deviation of 
${\bf T}$ is zero 
$$
\ddot{T^{\mu}}=-R^{\mu}_{\hspace{5 pt}
			\phi\rho\sigma}T^{\phi}T^{\rho}T^{\sigma}=0
$$
because the Riemann tensor is anti-symetric in the last two indeces.
Thus, this term gives no contribution to the deviation of
the volume, and the proof is unchanged.

\section{General Proof of the Scalar Curvature Relation}
\label{Scalarproof}
	It was stated in the text that the scalar curvature tells how a 
surface area of a generalized sphere in a curved space differs from the 
surface of a 
sphere in a flat space.  For our purposes a sphere of radius $r$ at a point 
$x_0$ is the set  of all points that are a geodesic distance $r$ from the
 given fixed point $x_0$.  In this section we will prove the relationship 
between curvature and boundary area by direct construction.

We begin by considering a point $x_0^{\mu}$, and finding the coordinates of
a point $x^{\mu}(\tau)$ a distance $\tau$ away in an arbitrary direction.  
Next we will consider how much these coordinates change when we change our
direction.  That is we will find $\frac {dx^{\mu}} {d\theta_i}$ where 
$\theta_i$ is a set of variables representing the change in direction.  
Using these changes in position we can find a small surface area element 
which we can integrate to find the total surface area of the sphere.      

Begin by considering the point $x_0^{\mu}$ and choosing a set of orthonormal
vectors $\{t_i^{\mu}\}$ at this point.  This is in essense equivalent to 
finding the Riemann Normal Coordinates at this point.  We will assume from 
now on that we are using Riemann Normal Coordinates around the point 
$x_0^{\mu}$.  We can now describe 
any unit vector $t^{\mu}$ at this
point as a linear superposition of these basis vectors.  If we use the 
standard polar angles in $D$-dimensions, then our arbitrary vector can
simply be thought of as the radial unit vector.  
For example, in two dimensions the radial vector is simply 
$$t^{\mu}(\theta)=t_1^{\mu}\cos{\theta}+t_2^{\mu}\sin{\theta}.$$

Now we want to find the coordinates of the point a small geodesic distance 
$\tau$ away from the point $x_0^{\mu}$ in the direction $t^{\mu}$
We start with the geodesic equation
$$
\frac {d^2 x^{\mu}} {d\tau^2} + \Gamma^{\mu}_{\hspace{5 pt} \nu \sigma}
	\frac {d x^{\nu}} {d\tau}\frac {d x^{\sigma}} {d\tau}=0
$$
and solve it iteratively to get $x^{\mu}(\tau)$ to order $\tau^3$: 
\begin{equation}
x^{\mu}(\tau)=x_0^{\mu}+\tau t^{\mu}
	-\frac {\tau^3} {6} \Gamma^{\mu}_{\hspace{5 pt} \nu \sigma, \alpha}
	t^{\nu}t^{\sigma}t^{\alpha}.
\end{equation}
It is important to note that because we are in Riemann Normal Coordinates, 
$\Gamma$ is 0 to 0th order in $\tau$.  That is why there is no term 
quadratic in $\tau$ and why the $\Gamma$ term involves a derivative. 
It is useful to symetrise the $\Gamma$ term in all three of the lower indices.
This can be done because it is already multiplying an object
that is symmetric in those indeces.  Thus we define
$$
\tilde{\Gamma}^{\mu}_{\hspace{5 pt}\nu \sigma, \alpha}=
\Gamma^{\mu}_{\hspace{5 pt}\nu \sigma, \alpha}
+\Gamma^{\mu}_{\hspace{5 pt}\alpha \nu, \sigma}
+\Gamma^{\mu}_{\hspace{5 pt}\sigma \alpha, \nu}
$$
so that we can write
$$x^{\mu}(\tau)=x_0^{\mu}+\tau t^{\mu}
	-\frac {\tau^3} {18}
	 \tilde{\Gamma}^{\mu}_{\hspace{5 pt}\nu \sigma, \alpha}
	t^{\nu}t^{\sigma}t^{\alpha}.$$

These are the coordinates of the point on our sphere in the direction 
$t^{\mu}$ from the starting point.  A small bit of surface area is then
simply
\begin{equation}
dA=\sqrt{det\biggl\{g(\frac {dx^\mu} {d\theta_i}, \frac {dx^\mu} {d\theta_j})\biggr\}}
d\theta_1 d\theta_2 \ldots d\theta_{D-1}.
\label{dAdef}
\end{equation}
$\frac {dx^\mu} {d\theta_i}$ is easily found to be
$$\frac {dx^\mu} {d\theta_i}=\tau \frac {dt^{\mu}} {d\theta_i}
	-\frac {\tau^3} {6} \tilde{\Gamma}^{\mu}_{\hspace{5 pt}
							\nu \sigma, \alpha}
	t^{\nu}t^{\sigma} \frac {dt^{\alpha}} {d\theta_i},$$
but we must be very careful in evaluating the inner product.  Namely it must
be evaluated not at $x_0$ but at $x(\tau)$.  To do this, we simply Taylor
expand the metric to order $\tau^2$ and find
$$
g_{\mu\nu}(x(\tau))=g_{\mu\nu}(x_0)+\frac {\tau^2} {2} g_{\mu\nu,\alpha\beta}
t^{\alpha}t^{\beta}.
$$

The inner product can now be evaluated to get
\begin{equation}
g(\frac {dx^\mu} {d\theta_i}, \frac {dx^\mu} {d\theta_j})=
g_{\mu,\nu} \tau^2 \frac {dt^\mu} {d\theta_i} \frac {dt^\nu} {d\theta_j}
+\frac {\tau^4} {2} (g_{\mu \nu, \alpha \beta} 
-\frac 2 3 \tilde{\Gamma}_{\mu \alpha \beta,\nu})
t^{\alpha}t^{\beta} 
\biggl\{ \frac {dt^\mu} {d\theta_i} \frac {dt^\nu} {d\theta_j} \biggr\},
\label{lengthbit}
\end{equation}
where 
$$\biggl\{ \frac {dt^\mu} {d\theta_i} \frac {dt^\nu} {d\theta_j} \biggr\}=
\frac 1 2 \biggl(
\frac {dt^\mu} {d\theta_i} \frac {dt^\nu} {d\theta_j}+
\frac {dt^\mu} {d\theta_j} \frac {dt^\nu} {d\theta_i} \biggr).
$$
After some mild algebra and a few rearrangements involving the symmetry of the
$i$ and $j$ terms we can show that the term in brackets is proportional to 
the Riemann tensor evaluated in Riemann Normal Coordinates:  

$$
g_{\mu\nu,\alpha\beta}-\frac 1 3 {\tilde{\Gamma}}_{\mu\beta\alpha,\nu}
-\frac 1 3 {\tilde{\Gamma}}_{\nu\beta\alpha,\mu}
=-\frac 1 3 (g_{\mu\alpha,\nu\beta}-g_{\mu\nu,\alpha\beta}
            -g_{\alpha\beta,\mu\nu}+g_{\nu\beta,\mu\alpha})
=-\frac 2 3R_{\mu \beta, \nu \alpha}.
$$
Equation \ref{lengthbit} then simplifies to
\begin{equation}
g(\frac {dx^\mu} {d\theta_i}, \frac {dx^\mu} {d\theta_j})=
g_{\mu,\nu} \tau^2 \frac {dt^\mu} {d\theta_i} \frac {dt^\nu} {d\theta_j}
-\frac {\tau^4} {3} (R_{\mu \beta, \nu \alpha})
t^{\alpha}t^{\beta} 
\biggl\{ \frac {dt^\mu} {d\theta_i} \frac {dt^\nu} {d\theta_j} \biggr\}.
\label{lengthbit2}
\end{equation}

The determinant in equation \ref{dAdef} is then
\begin{equation}
\det {g(\frac {dx^\mu} {d\theta_i}, \frac {dx^\mu} {d\theta_j})}=
\det \Biggl( {g_{\mu,\nu} \tau^2 \frac {dt^\mu} {d\theta_i} \frac {dt^\nu} {d\theta_j}
-\frac {\tau^4} {3} (R_{\mu \beta, \nu \alpha})
t^{\alpha}t^{\beta} 
\biggl\{ \frac {dt^\mu} {d\theta_i} \frac {dt^\nu} {d\theta_j} \biggr\}} \Biggr).
\label{det1}
\end{equation}

Before simplifying this determinant, we must make a brief digression and 
introduce some new structure.  Notice that 
$x^{\mu}=x_0^{\mu}+\tau t^{\mu}$ gives 
the coordinates of a point.  An 
equally good set of coordinates is $\Theta_I=(\tau,\theta_i)$.  If the $x^\mu$
represent the Riemann Normal Coordinates, the $\Theta_I$ represent the 
equivalent polar coordinates.  The metric
in this new set of coordinates is then
\begin{equation}
H_{IJ}=g_{\mu\nu} \frac {\partial x^{\mu}} {\partial \Theta_I} 
                  \frac {\partial x^{\nu}} {\partial \Theta_J}.
\end{equation}
From this we can easily find the inverse relationship
\begin{equation}
g^{\mu\nu}=H^{IJ} \frac {\partial x^{\mu}} {\partial \Theta_I} 
                  \frac {\partial x^{\nu}} {\partial \Theta_J}.
\label{gfromH}
\end{equation}

The various components of this matrix are
\begin{eqnarray}
H_{00}=g_{\mu\nu} t^{\mu} t^{\nu}=1;
 \\
H_{0i}=\tau g_{\mu\nu} t^{\mu} \frac {\partial t^{\nu}} {\partial \theta_j}
      = 0; \\
H_{ij}=\tau^2 g_{\mu\nu} \frac {\partial t^{\mu}} {\partial \theta_i} 
                         \frac {\partial t^{\nu}} {\partial \theta_j}      
      = \tau^2 h_{ij}.
\end{eqnarray} 
The first equation is simply the normalization of $t^{\mu}$.  The second 
equation being $0$ follows directly from this normalization.  If the 
magnitude of $\bf t$ is to remain unchanged it must be orthogonal to its 
deviation.  In the third equation, we have introduced a new matrix
$h_{ij}=g_{\mu\nu}\frac {\partial t^{\mu}} {\partial \theta_i} 
                         \frac {\partial t^{\nu}} {\partial \theta_j}$.

Because $H^{IJ}$ is block diagonal we can invert the blocks 
individually.
$$
H^{ij}=\frac 1 {\tau^2} h^{ij};
$$
$$
H^{00}=1.
$$
Substituting this into equation \ref{gfromH} we find that
\begin{equation}
g^{\mu\nu}=h^{ij} \frac {\partial t^{\mu}} {\partial \theta_i} 
                  \frac {\partial t^{\nu}} {\partial \theta_j}
          +t^{\mu} t^{\nu}.
\label{gfromh}
\end{equation}
This will be useful for simplifying our curvature equation.

Now we return to equation \ref{det1}.  
Notice that the first term is just $\tau^2 h_{ij}$ and the second term is 
smaller by a factor or $\tau^2$.  Using a standard perturbative expansion for
determinants we find that
$$
\det {g(\frac {dx^\mu} {d\theta_i}, \frac {dx^\mu} {d\theta_j})} =
\tau^{2(D-1)} \Bigl( \det{h_{ij}} \Bigr)
\biggl( 1 - \frac {\tau^2} {3} (R_{\mu \beta, \nu \alpha})
t^{\alpha}t^{\beta} 
\biggl\{ \frac {dt^\mu} {d\theta_i} \frac {dt^\nu} {d\theta_j} \biggr\} h^{ij} \biggr),
$$
where higher order terms have been dropped.  The first two factors are
just what we would get in flat space.  We can use them to define a flat space
area element 
$dA_{flat}=d\theta_1 \ldots d\theta_{D-1} \tau^{D-1} \sqrt{\det{h_{ij}}}$. 
The third factor shows that we now
have a small curvature dependent perturbation on the flat space result.
This perturbation can be simplified using equation \ref{gfromh}.
$$
(R_{\mu \beta, \nu \alpha})
t^{\alpha}t^{\beta} 
\biggl\{ \frac {dt^\mu} {d\theta_i} \frac {dt^\nu} {d\theta_j} \biggr\}
=(R_{\mu \beta, \nu \alpha})
t^{\alpha}t^{\beta} (g^{\mu\nu}-t^{\mu}t^{\nu})=
R_{\alpha\beta}t^{\alpha}t^{\beta}
$$
The $g^{\mu\nu}$ term causes the contraction, and the $t^{\mu}t^{\nu}$ term 
cancels due to the anti-symmetry of the Riemann tensor in the first two or 
last two indices.  

The boundary area of the sphere is then
\begin{equation}
A_{curved}=\int_{\rm All Angles} dA_{flat}
(1-\frac {\tau^2} {6}R_{\alpha\beta}t^{\alpha}t^{\beta})=
A_{flat}-\frac {\tau^2} 6 R_{\alpha\beta}\int_{\rm All Angles} dA_{flat}
	t^{\alpha}t^{\beta}.
\label{curvedA}
\end{equation}

We now need to integrate the 
term $t^{\mu}t^{\nu}$ over all angles.  Clearly, since $t^{\mu}$ is a 
radial unit vector, once integrated over all angles the term should be 
rotationally invariant.  Thus,
$$
\int_{\rm All Angles}t^{\mu}t^{\nu} dA_{flat} = a g^{\mu\nu}
$$
The proportionality constant can be found by contracting both sides and 
integrating,
$$
A_{flat}= a D
$$
so that
$$
a=\frac {A_{flat}}{D}.
$$

Equation \ref{curvedA} now simplifies to
$$
A_{curved}=A_{flat}(1-\frac {\tau^2} {6 D} R).
$$
We can then invert this to find
$$
R=\lim_{\tau \to 0}  \frac {6D} {\tau^2} (1-\frac {A_{curved}} {A_{flat}}). 
$$
Which completes the proof of equation \ref{scalardef1}.

\end{document}